\documentstyle[psfig]{mn}

\newif\ifAMStwofonts

\def\etal{{\it et\thinspace al.}\ }

\ifoldfss
  \ifCUPmtlplainloaded \else
    \NewTextAlphabet{textbfit} {cmbxti10} {}
    \NewTextAlphabet{textbfss} {cmssbx10} {}
    \NewMathAlphabet{mathbfit} {cmbxti10} {} 
    \NewMathAlphabet{mathbfss} {cmssbx10} {} 
  \fi
  \ifAMStwofonts
    \ifCUPmtlplainloaded \else
      \NewSymbolFont{upmath} {eurm10}
      \NewSymbolFont{AMSa} {msam10}
      \NewMathSymbol{\upi}     {0}{upmath}{19}
      \NewMathSymbol{\umu}     {0}{upmath}{16}
      \NewMathSymbol{\upartial}{0}{upmath}{40}
      \NewMathSymbol{\leqslant}{3}{AMSa}{36}
      \NewMathSymbol{\geqslant}{3}{AMSa}{3E}

      \let\leq=\leqslant 
      \let\geq=\geqslant 
    \fi
  \fi
\fi 

\ifnfssone
  \newmathalphabet{\mathit}
  \addtoversion{normal}{\mathit}{cmr}{m}{it}
  \addtoversion{bold}{\mathit}{cmr}{bx}{it}
  \newmathalphabet{\mathbfit} 
  \addtoversion{normal}{\mathbfit}{cmr}{bx}{it}
  \addtoversion{bold}{\mathbfit}{cmr}{bx}{it}
  \newmathalphabet{\mathbfss} 
  \addtoversion{normal}{\mathbfss}{cmss}{bx}{n}
  \addtoversion{bold}{\mathbfss}{cmss}{bx}{n}
  \ifAMStwofonts
    \ifCUPmtlplainloaded \else
      \UseAMStwoboldmath
      \makeatletter
      \new@mathgroup\upmath@group
      \define@mathgroup\mv@normal\upmath@group{eur}{m}{n}
      \define@mathgroup\mv@bold\upmath@group{eur}{b}{n}
      \edef\UPM{\hexnumber\upmath@group}
      \new@mathgroup\amsa@group
      \define@mathgroup\mv@normal\amsa@group{msa}{m}{n}
      \define@mathgroup\mv@bold\amsa@group{msa}{m}{n}
      \edef\AMSa{\hexnumber\amsa@group}
      \makeatother
      \mathchardef\upi="0\UPM19
      \mathchardef\umu="0\UPM16
      \mathchardef\upartial="0\UPM40
      \mathchardef\leqslant="3\AMSa36
      \mathchardef\geqslant="3\AMSa3E

      \let\leq=\leqslant 
      \let\geq=\geqslant 
    \fi
  \fi
\fi 

\ifnfsstwo
  \DeclareMathAlphabet{\mathbfit}{OT1}{cmr}{bx}{it}
  \SetMathAlphabet\mathbfit{bold}{OT1}{cmr}{bx}{it}
  \DeclareMathAlphabet{\mathbfss}{OT1}{cmss}{bx}{n}
  \SetMathAlphabet\mathbfss{bold}{OT1}{cmss}{bx}{n}
  \ifAMStwofonts
    \ifCUPmtlplainloaded \else
      \DeclareSymbolFont{UPM}{U}{eur}{m}{n}
      \SetSymbolFont{UPM}{bold}{U}{eur}{b}{n}
      \DeclareSymbolFont{AMSa}{U}{msa}{m}{n}
      \DeclareMathSymbol{\upi}{0}{UPM}{"19}
      \DeclareMathSymbol{\umu}{0}{UPM}{"16}
      \DeclareMathSymbol{\upartial}{0}{UPM}{"40}
      \DeclareMathSymbol{\leqslant}{3}{AMSa}{"36}
      \DeclareMathSymbol{\geqslant}{3}{AMSa}{"3E}

      \let\leq=\leqslant 
      \let\geq=\geqslant 
    \fi
  \fi
\fi 

\ifCUPmtlplainloaded \else
  \ifAMStwofonts \else 
    \def\upi{\pi}
    \def\umu{\mu}
    \def\upartial{\partial}
  \fi
\fi

\title{Photoionization and recombination of Fe~XIX}
\author[Hong Lin Zhang and Anil K. Pradhan]
       {Hong Lin Zhang$^1$ and Anil K. Pradhan$^2$ \\
       $^1$Applied Theoretical and Computational Physics Division,
Los Alamos National Laboratory, Los Alamos, NM 87545, USA\\
       $^2$Department of Astronomy, The Ohio State University, Columbus,
OH 43210, USA}
\date{Accepted  xxxxxx 
      Received xxxxxx;
      in original form xxxxxx}

\pagerange{\pageref{firstpage}--\pageref{lastpage}}
\pubyear{1994}

\def\LaTeX{L\kern-.36em\raise.3ex\hbox{a}\kern-.15em
    T\kern-.1667em\lower.7ex\hbox{E}\kern-.125emX}

\begin{document}

\maketitle

\label{firstpage}

\begin{abstract}
                Photoionization cross sections and recombination rate
coefficients are presented for the L-shell ground state fine structure levels
$2s^22p^4 \ ^3P_{2,0,1}$ of Fe~XIX. Several sets of calculations
including relativistic effects are
carried out: (i) Breit-Pauli R-matrix (BPRM), (ii) Relativistic
Distorted Wave (RDW), and (iii) a semi-relativistic calculation.
Non-relativistic LS coupling calculations are also done for comparison.
The BPRM calculations employ a configuration interaction target 
representation for Fe~XX consisting of 12 LS terms (23-fine structure
levels), as in the recently reported BPRM 
calculations by Donnelly \etal (MNRAS, 307, 595, 1999).
The background cross sections in all three sets of present calculations
agree with one another,
but differ considerably from those of Donnelly \etal. Owing to much more
extensive resonance structures in the present BPRM calculations, the sum of the
corresponding recombination rate coefficients for the $^3P_{2,0,1}$
levels are up to 50\% higher than the LS rates at low temperarures
but comparable for higher temperatures; in contrast
to the results of Donnelly \etal who obtained the LS rates to be higher
than their BPRM results
by about a factor of 2. Reasons for these discrepancies are discussed.
\end{abstract}


\section{INTRODUCTION}

 Photoionization and recombination are fundamental atomic processes in
astrophysical plasmas. However, rather elaborate and extensive atomic
calculations are required in order to compute the related
photoionization and recombination coefficients in an accurate and
self-consistent manner. In most existing astrophysical models these
atomic data are from methods that do not adequately consider the 
level of complexity attributable to the large number of
autoionizing resonances that manifest themselves in the photoionization
cross sections, and consequently, in the inverse process of (electron - ion) 
recombination. Most previous atomic calculations 
divide the recombination process into non-resonant `radiative recombination' 
(RR) and resonant `di-electronic recombination' (DR). This division is
valid only if either RR or DR is dominant in a given energy
region. In general, in the low energy region, the electron-ion recombination
may contain significant contributions from both the autoionizing 
resonances and the background cross section.
Therefore, an accurate theoretical treatment
requires both the RR and DR to be treated in a unified manner. 
Recent developments enable such a unified and self-consistent
calculation of photoionization and recombination cross sections, 
including highly detailed
resonance structures, using the close coupling approximation and the
R-matrix method (Nahar \& Pradhan 1992, 1994; NP1 and NP2). 
For highly charged ions, relativistic effects may
be considered in the Breit-Pauli R-matrix (BPRM) approximation 
(Zhang \& Pradhan 1997; Zhang, Nahar \& Pradhan (ZNP) 1999). 
Electron-ion recombination rate coefficients are thereby 
obtained including the RR and DR processes through ab initio calculations.

 In high temperature X-ray sources K- and L-shell
atomic ions are prominent constituents. In particular the highly ionized
ions of iron are important in laboratory and astrophysical sources
(Liedahl \etal 1992). 
Recently, Donnelly \etal (1999) employed
the BPRM approximation to calculate total and partial 
photoionization cross sections for
the ground state fine structure levels $2s^22p^4 \ ^3P_{2,0,1}$
of Fe~XIX. They also calculated recombination rate
coefficients for the transitions from the ground level $2s^22p^3\ ^4S^o_{3/2}$
in Fe~XX to those levels in Fe~XIX using the partial photoionization
cross sections. They employed a
sophisticated representation for the 12 lowest SL$\pi$ terms
corresponding to 23 $J\pi$ fine structure levels in the target ion
Fe~XX. Donnelly \etal obtained fine structure photoionization cross
sections for the three ground state levels such that the individual
cross sections differed considerably from each other, in background and
in details of resonances, and the 
weighted average differed from the non-relativistic LS values by 
more than a factor of two. The results were surprising since they imply
relativistic effects are larger than expected. Otherwise, the individual 
fine structure cross sections should have a background of the same
magnitude (although different resonance structures). 
For example, an earlier BPRM study on C-like Fe~XXI, with
the same ground state symmetries $^3P_{0,1,2}$ showed that 
the background cross sections for the 
three levels were nearly equal (Zhang 1998). 
We have therefore repeated the
calculations of Donnelly \etal as precisely as possible, using the same
eigenfunction expansion as in their work. We obtain very different
results, confirmed by several other calculations in different 
approximations, as well as much smaller relativistic effects.
New recombination rate coefficients for the Fe~XIX ground levels are
also presented and compared.

\section{THEORY}
The extension of the close coupling method to electron-ion recombination
is described in earlier works (NP1, NP2, ZNP), together with the
details of the unified treatment. Here we present a brief description of
the theory relevant to the present calculations.
The calculations are carried out in the close coupling (CC) approximation
employing the $R$-matrix method, in LS coupling and in intermediate
coupling with the BP Hamiltonian. The target ion
is represented by an $N$-electron system, and the total wavefunction
expansion,
$\Psi(E)$, of the ($N$+1) electron-ion system of symmetry $SL\pi$ (LS
coupling) or
$J\pi$ (BP, intermediate coupling) may be
represented in terms of the target eigenfunctions as:

\begin{equation}
\Psi(E) = A \sum_{i} \chi_{i}\theta_{i} + \sum_{j} c_{j} \Phi_{j},
\end{equation}

\noindent
where $\chi_{i}$ is the target wavefunction in a specific state
$S_iL_i\pi_i$ or $J_i\pi_i$ and $\theta_{i}$ is the wavefunction for the
($N$+1)-th electron in a channel labeled as
$S_iL_i(J_i)\pi_ik_{i}^{2}\ell_i(SL\pi \ {\rm or}\ J\pi)$;
$k_{i}^{2}$
being its incident kinetic energy. $\Phi_j$'s are the correlation
functions of the ($N$+1)-electron system that account for short range
correlation and the orthogonality between the continuum and the bound
orbitals.  Most of the present calculations are in the BP approximation
(Pradhan \& Zhang 1997, Zhang 1998).

Recombination of an incoming electron to the target ion may occur
through non-resonant, background continuum, usually referred to as
radiative recombination (RR):

\begin{equation}
e + X^{++} \rightarrow  h\nu + X^+,
\end{equation}

\noindent
which is the inverse process of direct photoionization, or through the
two-step recombination process via autoionizing resonances, i.e.
dielectronic recombination (DR):
\begin{equation}
e + X^{++} \rightarrow (X^+)^{**}  \rightarrow  \left\{
\begin{array}{c}  e + X^{++} \\   h\nu + X^+ \end{array} \right. ,
\end{equation}

\noindent
where the incident electron is in a
quasi-bound doubly-excited state which leads either to (i)
autoionization,
a radiation-less transition to a lower state  of the ion and the free
electron, or to (ii) radiative stabilization predominantly
via decay of the ion core,
usually to the ground state, and the bound electron.

In the unified treatment the photoionization cross sections,
$\sigma_{\rm PI}$,
of a large number of low-$n$ bound states -- all possible states with
$n \leq n_{\rm max}\sim 10$ -- are obtained in the close coupling (CC)
approximation
as in the Opacity Project (Seaton 1987). Coupled channel calculations
for
$\sigma_{\rm PI}$ include both
the background and the resonance structures (due to the doubly excited
autoionizing states) in the cross sections. The recombination cross
section, $\sigma_{\rm RC}$, is related to $\sigma_{\rm PI}$, 
through detailed balance (Milne relation) as

\begin{equation}
\sigma_{\rm RC}(\epsilon) =
{\alpha^2 \over 4} {g_j\over g_i}{(\epsilon + I_{\rm p})^2\over
\epsilon}\sigma_{\rm PI}
\end{equation}
in Rydberg units; 
$g_i$ and $g_j$ are the statistical weight factors of the recombining 
(target or residual) and recombined ions respectively;
$\alpha$ is the fine structure constant, $\epsilon$ is
the photoelectron energy, and $I_{\rm p}$ is the ionization potential. In the
present work, it is assumed that the recombining ion is in the ground
state, and recombination can take place into the ground or any of the
excited recombined (e+ion) states. The contributions of these bound
states to the total $\sigma_{\rm RC}$ are obtained by summing over the
contributions
from individual cross sections. $\sigma_{\rm RC}$ thus obtained from
$\sigma_{\rm PI}$, including the autoionizing resonances, corresponds to
the
total (DR+RR) unified recombination cross section.

Recombination into the high-$n$ states,
i.e. $n_{\rm max} < n \leq \infty$, 
must also be included (Fig.~1 of NP1).
To each excited threshold $S_iL_i(J_i)\pi_i$
of the $N$-electron target ion, there corresponds an infinite series of
($N$+1)-electron states, $S_iL_i(J_i)\pi_i\nu\ell$,
to which recombination can occur, where $\nu$ is the effective
quantum number. For these states
DR dominates the recombination process and the background recombination
is negligibly small. The contributions from these
states are added by calculating the collision strengths, $\Omega_{\rm DR}$,
employing the precise theory of radiation damping by Bell \& Seaton (1985). 
Several aspects related to the application of the theory to
the calculation of DR collision strengths are described in the
references cited.

 The total (e + ion) recombination rate coefficients may be calculated
using the photoionization cross sections $\sigma_{\rm PI}(\epsilon)$ and the
DR collision strengths $\Omega_{\rm DR}$. In the present work however, we
confine ourselves to a calculation of the level-specific 
$\sigma_{\rm RC}(T,i \rightarrow j)$, 
where $j$ refer to the ground state fine structure levels
$^3P_{2,0,1}$. The corresponding maxellian averaged rate coefficients 
are obtained as

\begin{eqnarray}
\alpha_{\rm RC}(T;i \rightarrow j) = & {\textstyle g_j \over \textstyle g_i} 
{\textstyle 2 \over \textstyle kTc\sqrt{2\pi m^3kT}} \cr
\times & \int_0^\infty {E^2\sigma_{\rm PI} (\epsilon;j \rightarrow i)
e^{-\epsilon/kT}d\epsilon},
\end{eqnarray}
where $E = h\nu = \epsilon + I_{\rm p}$ is the photon energy.

\section{COMPUTATIONS}
\subsection{Residual ion representation for Fe~XX}

 Close coupling calculations for an (e + ion) system 
require an accurate wavefunction expansion for the states of interest
in the target or the residual ion. Donnelly \etal (1999) used an
elaborate CI representation for the Fe~XX states obtained using the
atomic structure code CIV3 (Hibbert 1975). We obtain a very similar CI
target expansion using  the SUPERSTRUCTURE  program (Eissner \etal
1974), with the same target states as Donnelly \etal, i.e. the ones
dominated by configurations $2s^22p^3, 2s2p^4, 2p^5, 2s^22p^23s$ and all
associated terms and fine structure levels.
Both the CIV3
and the SUPERSTRUCTURE are carried out in intermediate coupling using
the Breit-Pauli approximation. Table~1 
presents the fine-structure level energies from the present BPRM calculation, 
which are almost identical to those from the SUPERSTRUCTURE 
results, together with experimental energies (Sugar \& Corliss 1985)
and those calculated by Donnelly \etal using CIV3.
The present calculated energies in Table 1 are taken after intermediate
coupling stage for the target ion in the code RECUPD of the BPRM package
of codes (Scott and Taylor 1982, Hummer \etal 1993, Berrington \etal
1995).
As is readily seen, the agreement
between the calculated and observed energies is very good, better than
1\% for most of the levels (the present calculated energies are slightly
better in agreement with experiment than those of Donnelly \etal)
Table 2 presents a subset of the oscillator strengths calculated, for
the transition array $2s^22p^3  \longrightarrow 2s2p^4$, and
compared with those of Donnelly \etal, and from
the U.S. National Institute for Standards
and Technology (NIST) compilation by
Fuhr \etal (1988) based on the relativistic calculations by Cheng \etal
(1979). All three sets of oscillator strengths are in good
agreement; the present values are in somewhat better agreement with the
NIST tabulation than Donnelly \etal for some transitions.
We may therefore deduce that the wavefunction representations in the two
calculations, Donnelly \etal and ours, are of the same accuracy.

\begin{table}
\begin{minipage}{80mm}
\caption{Calculated (cal) LSJ energy levels of Fe~XX in Rydbergs
and comparison with the observed (obs) energies
from NIST (1985), and with those (DBK) by Donnelly \etal (1999).}
\begin{tabular}{rlllrrr}
\hline
$i$  & Level      &  &   $E_{cal}$(LSJ) &  $E_{obs}$ & $E$(DBK) \\ 
\hline
 1& $2s^22p^3 \ ^4S^o$  &   $ 3/2 $&  0.0000    & 0.0000 & 0.0000   \\
 2& $2s^22p^3 \ ^2D^o$  &   $ 3/2 $&  1.2789    & 1.2632 & 1.2887   \\
 3& $2s^22p^3 \ ^2D^o$  &   $ 5/2 $&  1.6446    & 1.6050 & 1.6612   \\
 4& $2s^22p^3 \ ^2P^o$  &   $ 1/2 $&  2.3700    & 2.3718 & 2.3881   \\
 5& $2s^22p^3 \ ^2P^o$  &   $ 3/2 $&  2.9332    & 2.9465 & 2.9616   \\
 6& $2s2p^4   \ ^4P$  &   $ 5/2 $&  6.8499    & 6.8594 & 6.8674   \\
 7& $2s2p^4   \ ^4P$  &   $ 3/2 $&  7.4451    & 7.4799 & 7.4657   \\
 8& $2s2p^4   \ ^4P$  &   $ 1/2 $&  7.6432    & 7.6796 & 7.6629   \\
 9& $2s2p^4   \ ^2D$  &   $ 3/2 $&  9.5480    & 9.5006 & 9.5729   \\
 10& $2s2p^4  \ ^2D$  &   $ 5/2 $&  9.6762    & 9.6445 & 9.7304   \\
 11& $2s2p^4  \ ^2S$  &   $ 1/2 $&  10.9215    & 10.8920 & 10.9652   \\
 12& $2s2p^4  \ ^2P$  &   $ 3/2 $&  11.3899    & 11.3219 & 11.4352   \\
 13& $2s2p^4  \ ^2P$  &   $ 1/2 $&  12.2595    & 12.2113 & 12.3052   \\
 14& $2p^5    \ ^2P^o$  &   $ 3/2 $&  17.8936    & 17.8109 & 17.9667   \\
 15& $2p^5    \ ^2P^o$  &   $ 1/2 $&  18.8721    & 18.7922 & 18.9503   \\
 16& $2s^22p^23s \ ^4P$  &   $ 1/2 $&  65.8494    & --  & 65.3888 \\
 17& $2s^22p^23s \ ^4P$  &   $ 3/2 $&  66.3955    & -- & 65.9709 \\
 18& $2s^22p^23s \ ^4P$  &   $ 5/2 $&  66.7934    & -- & 66.3752  \\
 19& $2s^22p^23s \ ^2P$  &   $ 1/2 $&  66.6709    & -- & 66.2395  \\
 20& $2s^22p^23s \ ^2P$  &   $ 3/2 $&  67.0759    & -- & 66.4774   \\
 21& $2s^22p^23s \ ^2D$  &   $ 5/2 $&  67.9909    & -- & 67.5913   \\
 22& $2s^22p^23s \ ^2D$  &   $ 3/2 $&  68.0771    & -- & 67.6739   \\
 23& $2s^22p^23s \ ^2S$  &   $ 1/2 $&  69.4502    & -- & 68.7533   \\
\hline
\end{tabular}
\end{minipage}
\end{table}

\begin{table}
\begin{minipage}{80mm}
\caption[ ]{Dipole oscillator strengths ( $\geq$ 0.0001) for fine 
structure transitions
in the transition array $2s^22p^3 \rightarrow 2s2p^4$ for the target ion
Fe~XX; comparisons are
with the NIST data by Cheng \etal (in Fuhr \etal 1988), and with 
those (DBK) by Donnelly \etal (1999).}
\begin{tabular}{lllllll}
\hline
\multicolumn{3}{c}{Transition}  & Present & NIST &  DBK \\ 
\hline
$^4S^o_{3/2}$& -- &$^4P_{1/2} $ &  0.0208   & 0.0221 & 0.0208   \\
             & -- &$^4P_{3/2} $ &  0.0392   & 0.0413 & 0.0391  \\
             & -- &$^4P_{5/2} $ &  0.0503   & 0.0520 & 0.0485  \\
             & -- &$^2S_{1/2} $ &  0.0009   & 0.0010 & 0.0011  \\
             & -- &$^2P_{3/2} $ &  0.0040   & 0.0045 & 0.0045  \\
             & -- &$^2D_{3/2} $ &  0.0021   & 0.0026 & 0.0028  \\
$^2P^o_{1/2}$& -- &$^4P_{1/2} $ &  0.0010   & 0.0011 & 0.0012  \\
             & -- &$^2S_{1/2} $ &  0.0622   & 0.0640 & 0.0615  \\
             & -- &$^2P_{1/2} $ &  0.0057   & 0.0057 & 0.0045  \\
             & -- &$^2P_{3/2} $ &  0.0272   & 0.0284 & 0.0279  \\
             & -- &$^2D_{3/2} $ &  0.0141   & 0.0146 & 0.0130  \\
$^2P^o_{3/2}$& -- &$^4P_{3/2} $ &  0.0010   & 0.0011 & 0.0011  \\
             & -- &$^4P_{5/2} $ &  0.0003   & 0.0003 & 0.0003  \\
             & -- &$^2S_{1/2} $ &  0.0026   & 0.0030 & 0.0018  \\
             & -- &$^2P_{1/2} $ &  0.0679   & 0.0700 & 0.0677  \\
             & -- &$^2P_{3/2} $ &  0.0169   & 0.0167 & 0.0162  \\
             & -- &$^2D_{3/2} $ &  0.0018   & 0.0020 & 0.0016  \\
             & -- &$^2D_{5/2} $ &  0.0242   & 0.0250 & 0.0236  \\
$^2D^o_{3/2}$& -- &$^4P_{1/2} $ &  0.0004   & 0.0005 & 0.0005  \\
             & -- &$^4P_{3/2} $ &  0.0003   & 0.0004 & 0.0003  \\
             & -- &$^4P_{5/2} $ &  0.0033   & 0.0038 & 0.0041  \\
             & -- &$^2S_{1/2} $ &  0.0289   & 0.0300 & 0.0297  \\
             & -- &$^2P_{1/2} $ &  0.0148   & 0.0151 & 0.0132  \\
             & -- &$^2P_{3/2} $ &  0.0187   & 0.0181 & 0.0159  \\
             & -- &$^2D_{3/2} $ &  0.0742   & 0.0780 & 0.0745  \\
$^2D^o_{5/2}$& -- &$^4P_{3/2} $ &  0.0001   & 0.0001 & 0.0002  \\
             & -- &$^4P_{5/2} $ &  0.0011   & 0.0012 & 0.0013  \\
             & -- &$^2P_{3/2} $ &  0.0868   & 0.0890 & 0.0853  \\
             & -- &$^2D_{5/2} $ &  0.0599   & 0.0630 & 0.0589  \\
\hline
\end{tabular}
\end{minipage}
\end{table}

\subsection{Photoionization}

 The BPRM method as applied to self-consistent photoionization and recombination
calculations has been discussed in detail in several previous works
and references therein (e.g. ZNP, Pradhan \& Zhang 1998). 
The use of such data to astrophysical applications in photoionization modeling, 
and photoionization and recombination data for all ions of carbon, nitrogen,
and oxygen are described by Nahar \& Pradhan (1998) and Nahar (1999).
 Donnelly \etal also provide a detailed discussion of 
the R-matrix method and the calculations. In the present work we
confine the calculations to the ground term levels considered by
Donnelly \etal for both the total and partial photoionization cross
sections. 

In addition to the BPRM calculations using the target expansion described
in the above sub-section, non-relativistic LS coupling
calculations, using the same target, were carried out for comparison. 
We also obtained two other data sets for the photoionization
of Fe~XIX, with the relativistic distorted wave (RDW) code (Zhang 1998),
and with the semi-relativistic atomic structure code CATS (Abdallah \etal 1988)
and ionization code GIPPER (Clark \& Abdallah, unpublished) of the Los Alamos
National Laboratory (LANL).

\subsection{Recombination rate coefficients}

Maxwellian averaged recombination rate coefficients for the ground term
levels $2s^22p^4 \ ^3P_{2,0,1}$ are
obtained using the partial photoionization cross sections $\sigma_{PI}$
for the transitions
$h\nu + 2s^22p^4 \ (^3P_J) \longrightarrow e + 2s^22p^3 (^4S^o_{3/2})$.
The recombination rates were also calculated using the LS coupling
photoionization cross sections for $h\nu + 2s^22p^4 \ (^3P) \longrightarrow
e + 2s^22p^3 (^4S^o)$, for comparison with the relativistic calculations.

\subsection{Oscillator strengths}

 Another indication of the accuracy of the bound-free photoionization
calculations may be obtained by calculating the bound-bound radiative
transition probabilities using the BPRM method for the fine structure
E1 (dipole allowed and intercombination) transitions 
(e.g. Nahar and Pradhan 1999a,b). As the same bound state wavefunction
expansions are employed in both the bound-free and the bound-bound
transitions, one expects the same accuracy. A number of oscillator
strengths of Fe~XIX were computed and compared with other accurate
calculations using the multiconfiguration Dirac-Fock (MCDF) method by
Cheng \etal (1979).

\section{RESULTS AND DISCUSSION}

Results for photoionization, recombination, and transition
probabilities calculated in the present work
are consistent with one another, as discussed below.

\subsection{Photoionization cross sections}

Fig.~1 presents the total photoionization cross sections for the three
ground term fine structure levels $^3P_{2,0,1}$ 
using the BPRM method (the solid lines) and the RDW method (the dashed lines).
The non-relativistic
LS coupling photoionization cross sections of the $^3P$ term and 
the weighted average of the RDW results are also presented (Fig.~1d).
Comparing our Fig.~1 with Figs.~1 and 2 in Donnelly \etal one sees
the present BPRM results differ considerably with those
by Donnelly \etal in the following three aspects:
1) while the results of Donnelly \etal exhibit large variations 
in the background cross sections for the three levels, 
the present BPRM results show those to be nearly the same;
2) in the photon energy range $\sim 126 - 154$ Ry, the cross sections for
levels $^3P_2$ and $^3P_1$ in Donnely \etal's Fig.~1 show extensive and
strong resonance structures while our results do not;
3) the weighted average (not shown directly here) 
of the background cross sections for the three fine structure
levels in our results would be very close to the non-relativistic
LS results, as is obvious from Fig.~1, while those in
Donnelly \etal differ from the non-relativistic LS values by
more than a factor of two. 

However, the detailed resonance structures in
the present BPRM calculations are more extensive and differ
significantly from the LS calculations. Fig.~2 
shows the cross sections at higher resolution. It is apparent that the
resonance positions in the LS $^3P$ results (panel d) do not correspond
in detail or magnitude to the individual resonances in the cross
sections for the fine structure components $^3P_{2,0,1}$ (panels a,b,c
respectively).

Fig.~3 presents the partial cross sections for photoionization from the
ground level $^3P_2$ of Fe~XIX into a few fine structure levels
of the residual ion Fe~XX. Comparing this figure with Fig.~3 in
Donnelly \etal, one again sees that similar significant differences exist 
in both the background and the resonances, especially the resonances
in the photoelectron energy region 20 -- 45 Ry.

Now we discuss the above three major differences between the present
results and those in Donnelly \etal
1) From our many other calculations, such as those shown in Fig.~1 
for Fe~XXII and Figs.~(4c) and (4d) for Fe~XXIV in Zhang (1998),
the relativistic effects do not appear to
affect the background photoionization cross sections
very much, and the backgrounds for the fine structure
levels corresponding to the same energy term appear to be
almost the same. Those BPRM results, as well as the
present results, were all confirmed by the RDW calculations
(shown as dotted or dashed lines in the corresponding figures). 
The backgrounds in the present BPRM results also agree
with the semi-relativistic results from the LANL codes
(basically similar to the RDW results and therefore not shown).
2) The resonances in the photon energy region $\sim 126 - 154$ Ry,
or the photoelectron energy region $\sim 20 - 45$ Ry, obtained by
Donnelly \etal would correspond to the process
\begin{equation}
{\rm h}\nu + 2s^22p^4 \rightarrow 2\ell^4 3\ell n'\ell
\rightarrow 2s^22p^3 + e,
\end{equation}
with $n' \geq 3$;
it is obvious that the first step in this process involves two-electron jumps
and is not allowed (there would be some contributions through the configuration
mixing among the $n=2$ and 3 states but these would not produce such 
strong resonances).
3) As is clear from the discussion in 1), the background in the
weighted average of the results for 
the three levels should not differ much from the
non-relativistic LS-coupling results. This should be true
for ions with low and intermediate nuclear charge $Z$, where
the relativistic effect should not be significant enough to affect
the background. Indeed, we usually
use this fact to check our calculations.

 The detailed resonance structures on the other hand may differ
considerably since they correspond to different target thresholds in the
BPRM and the LS coupling calculations. In the relativistic calculations
the Rydberg series of resonances are more extensive owing to the 
many more fine structure
level splittings in the target than the LS terms.

 We also note that a preliminary set of BPRM calculation with a smaller
number of target terms, excluding the $2s^22p^23s$, also yielded
essentially the same results for the photoionization cross sections.
This is expected for highly charged ions since the configuration
interaction between different n-complexes decreases with ion charge.

\subsection{Recombination rate coefficients}

\begin{table}
\centering
\begin{minipage}{80mm}
\caption[ ]{The present BPRM recombination rates 
as functions of temperature for the
transitions from the ground level $2s^22p^3\ ^4S^o_{3/2}$ in Fe~XX 
to the three levels of Fe~XIX in its ground term $2s^22p^4\ ^3P_J$; the sum
of the values for these levels (SUM) is also shown and compared with the 
non-relativistic LS-coupling results (LS).}
\begin{flushleft}
\begin{tabular}{cccccc}
\hline
$T$(K) &    $^3P_2$ &    $^3P_0$ &   $^3P_1$  &   SUM      &    LS     \\
\hline
5.0[4]&   2.57[-11]&   4.24[-12]&   1.11[-11]&   4.11[-11]&   2.78[-11] \\
6.0[4]&   2.31[-11]&   3.99[-12]&   1.03[-11]&   3.74[-11]&   2.57[-11] \\
7.0[4]&   2.12[-11]&   3.77[-12]&   9.63[-12]&   3.46[-11]&   2.39[-11] \\
8.0[4]&   1.96[-11]&   3.55[-12]&   9.06[-12]&   3.22[-11]&   2.24[-11] \\
9.0[4]&   1.83[-11]&   3.36[-12]&   8.55[-12]&   3.02[-11]&   2.10[-11] \\
1.0[5]&   1.71[-11]&   3.18[-12]&   8.09[-12]&   2.84[-11]&   1.98[-11] \\
2.0[5]&   1.04[-11]&   2.01[-12]&   5.15[-12]&   1.75[-11]&   1.25[-11] \\
3.0[5]&   7.44[-12]&   1.45[-12]&   3.75[-12]&   1.26[-11]&   9.15[-12] \\
4.0[5]&   5.81[-12]&   1.13[-12]&   2.94[-12]&   9.88[-12]&   7.24[-12] \\
5.0[5]&   4.77[-12]&   9.33[-13]&   2.43[-12]&   8.12[-12]&   6.01[-12] \\
6.0[5]&   4.05[-12]&   7.93[-13]&   2.07[-12]&   6.91[-12]&   5.16[-12] \\
7.0[5]&   3.52[-12]&   6.90[-13]&   1.80[-12]&   6.01[-12]&   4.54[-12] \\
8.0[5]&   3.12[-12]&   6.12[-13]&   1.60[-12]&   5.33[-12]&   4.06[-12] \\
9.0[5]&   2.80[-12]&   5.50[-13]&   1.43[-12]&   4.79[-12]&   3.68[-12] \\
1.0[6]&   2.55[-12]&   5.01[-13]&   1.30[-12]&   4.35[-12]&   3.38[-12] \\
2.0[6]&   1.38[-12]&   2.70[-13]&   6.98[-13]&   2.34[-12]&   1.94[-12] \\
3.0[6]&   9.63[-13]&   1.89[-13]&   4.86[-13]&   1.64[-12]&   1.41[-12] \\
4.0[6]&   7.43[-13]&   1.45[-13]&   3.73[-13]&   1.26[-12]&   1.11[-12] \\
5.0[6]&   6.03[-13]&   1.18[-13]&   3.02[-13]&   1.02[-12]&   9.12[-13] \\
6.0[6]&   5.04[-13]&   9.84[-14]&   2.52[-13]&   8.54[-13]&   7.70[-13] \\
7.0[6]&   4.30[-13]&   8.40[-14]&   2.15[-13]&   7.29[-13]&   6.62[-13] \\
8.0[6]&   3.73[-13]&   7.29[-14]&   1.86[-13]&   6.32[-13]&   5.78[-13] \\
9.0[6]&   3.28[-13]&   6.40[-14]&   1.63[-13]&   5.55[-13]&   5.10[-13] \\
1.0[7]&   2.91[-13]&   5.69[-14]&   1.45[-13]&   4.93[-13]&   4.54[-13] \\
\hline
\end{tabular}
\end{flushleft}
\end{minipage}
\end{table}

 Table~3 presents the $\alpha_{\rm R}(T)$ calculated using 
the $\sigma_{\rm RC}$
obtained from the partial photoionization cross sections $\sigma_{\rm PI}
(^3P_J \rightarrow\ ^4S^o_{3/2})$, for the three levels $J=2,0,1$.
The sum of the recombination rate coefficients for
these three fine structure levels is also given, together with
the LS coupling results. It is important to note that the total
recombination rate coefficient into a LS term is a {\it simple sum} of 
the values for the individual fine
structure levels corresponding to this term. 
This is because the expression for the
recombination rate (see our Eq.~(5) or Eq.~(7) in Donnelly \etal)
already includes a multiplication by the statistical weight factor, 
which is $g_j=(2J + 1)$ in our Eq.~(5) and $\omega_j$ in
Donnelly \etal's Eq.~(7).
We note that, as shown by Eq.~(3) of NP2, the {\it total} recombination rate 
coefficient for a given target level
is obtained by {\it summing} over contributions 
from all possible states or levels in the recombined ion.
Donnelly \etal, on the other hand,
performed a ``weighted average'' in order to obtain the
recombination rate coefficient for the LS term -- 
leading to a further error in their
numerical values -- and found a different kind of discrepancy
compared with the LS coupling results. 

The present rate coefficients for the $^3P$ term obtained by summing
over the fine structure values are also significantly higher than the
LS coupling results, especially for the low temperatures, the BPRM
results being larger by up to 50\%.
In contrast, the BPRM results in Donnelly \etal are up to a factor of 2
smaller than the LS values.
The differences between the BPRM and the LS
coupling values in the present results are due to the different 
details of the resonance structures
in the two cases. Whereas in LS coupling the
fine structure thresholds belonging to a given LS term are degenerate, 
these are explicitly included in the BPRM calculations at the appropriate
energies and consequently the resonances are more numerous. 
As displayed earlier in Fig.~2, the resonance positions and shapes are also 
different in the BPRM and LS coupling calculations.
Further, more levels of the autoionizing terms which do not autoionize in
the LS coupling case contribute in the BPRM case (for example, see
the discussion on the $1s2p3p$ autoionizing states in Sec.~IIIB
in Zhang (1998)).
At higher temperatures the differences between
the present LS coupling and BPRM rates are smaller, since the overall
background cross sections are of similar magnitude at higher energies, 
above all $n=2$ target thresholds. It might also be pointed out that
the autoionizing resonances are unlikely to be significantly dampened
due to radiative decays, to affect the overall cross sections and rates, 
since they involve $\Delta \ n = 0$ transitions with much
smaller radiative rates than autoionization rates. Further discussion
of recombination theory in the close coupling formulation, including the
R-matrix method, is 
given in several previous works: Davies and
Seaton (1969), Bell and Seaton (1985), Robicheaux \etal (1995),
Gorczyca and Badnell (1997), Pradhan and Zhang (1997), Badnell
\etal (1998), Robicheaux (1998), Seaton (1998), and Zhang \etal (1999).

\subsection{Transition Probabilities}
 
 Transition probabilities provide an independent test of the accuracy of
the bound state wavefunctions.
Fine structure transition probabilities may be computed for E1 dipole
allowed and intercombination transitions using recent developoments in
the BPRM method (Hummer \etal 1993, Berrington \etal 1998,
Nahar and Pradhan, 1999a,b). We calculate the oscillator
strengths for the fine structure components within several multiplets
corresponding to the initial $^3P$ symmetry and final $^{3,1}(S,P,D)^o$
symmetries. Table~4 presents a subset of these results: the gf-values 
for the transitions
$2s^22p^4 \ ^3P_{2,0,1} \longrightarrow 2s2p^5 \ ^{3,1}P^o_{1,2}$. 
The present results are
compared with the MCDF calculations by Cheng \etal (1979) that have
been incorporated in the evaluated compilation by the U.S. National Institute of
Standards and Technology (NIST; Fuhr \etal 1988). In all cases  the
agreement between the present BPRM results and the MCDF values is within
about 10\%. 
Both the length and the velocity oscillator strengths are
obtained, and found to be in good agreement for the stronger transitions
(gf $>$ 0.01) in all multiplets calculated.

\begin{table}
\begin{minipage}{80mm}
\caption{Dipole oscillator strengths for fine structure transitions
in the transition array $2s^22p^4\ ^3P \rightarrow 2s2p^5\ ^{3,1}P^o$ 
for Fe~XIX; comparisons are
with the NIST data by Cheng \etal (in Fuhr \etal 1988}
\begin{tabular}{@{}llllll}
\hline
&\multicolumn{3}{c}{Transition}        & Present &  NIST  \\
\hline
& $^3P_2$ & -- & $^3P^o_1$ & 0.129  &  0.147  \\
&         & -- & $^1P^o_1$ & 0.0313 &  0.0355 \\
&         & -- & $^3P^o_2$ & 0.296  &  0.34   \\
& $^3P_0$ & -- & $^3P^o_1$ & 0.0772 &  0.087  \\
&         & -- & $^1P^o_1$ & 0.0037 &  0.005  \\
& $^3P_1$ & -- & $^3P^o_0$ & 0.0898 &  0.1035 \\
&         & -- & $^3P^o_1$ & 0.0617 &  0.0705 \\
&         & -- & $^1P^o_1$ & 0.0029 &  0.003  \\
&         & -- & $^3P^o_2$ & 0.0986 &  0.1122 \\
\hline
\end{tabular}
\end{minipage}
\end{table}

\section*{Acknowledgments}
The work by HLZ was performed under the auspices of the US Department of Energy.
The work by AKP was partially supported by grants from the US National
Science Foundation (AST-9870089) and NASA (NAG5 6908). 
The computational work was carried out
at the Ohio Supercomputer Center in Columbus Ohio.

\def\amp{{\it Adv. At. Molec. Phys.}\ }
\def\apj{{\it Astrophys. J.}\ }
\def\apjs{{\it Astrophys. J. Suppl. Ser.}\ }
\def\apjl{{\it Astrophys. J. (Letters)}\ }
\def\aj{{\it Astron. J.}\ }
\def\aa{{\it Astron. Astrophys.}\ }
\def\aasup{{\it Astron. Astrophys. Suppl.}\ }
\def\adndt{{\it At. Data Nucl. Data Tables}\ }
\def\cpc{{\it Comput. Phys. Commun.}\ }
\def\jqsrt{{\it J. Quant. Spectrosc. Radiat. Transfer}\ }
\def\jpb{{\it Journal Of Physics B}\ }
\def\pasp{{\it Pub. Astron. Soc. Pacific}\ }
\def\mn{{\it MNRAS}\ }
\def\psc{{\it Physica Scripta}\ }
\def\pra{{\it Physical Review A}\ }
\def\prl{{\it Physical Review Letters}\ }
\def\zpds{{\it Z. Phys. D Suppl.}\ }

\clearpage

\begin{figure}
\centering
\psfig{figure=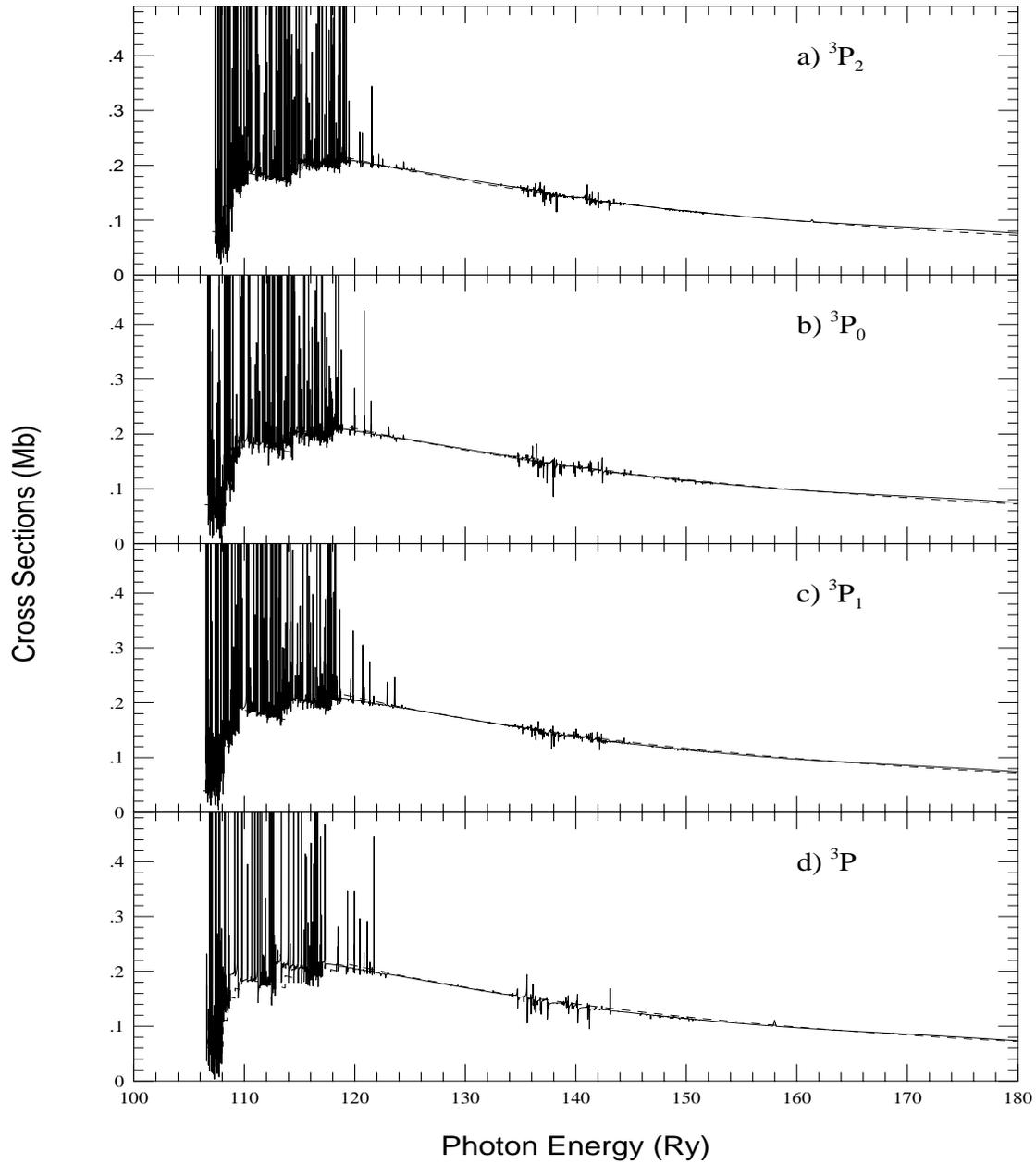,height=20.0cm,width=18.0cm}
\caption{Total photoionization cross sections for the three levels
$J=2,0,1$ (a, b and c) of the ground term $2s^22p^4\ ^3P$ in Fe~XIX
by the present
BPRM (solid lines) and the RDW (dashed lines) calculations.
The non-relativistic LS coupling results are also presented (d), together
with the average values of the RDW results (the dashed line).}
\end{figure}

\clearpage
\begin{figure}
\centering
\psfig{figure=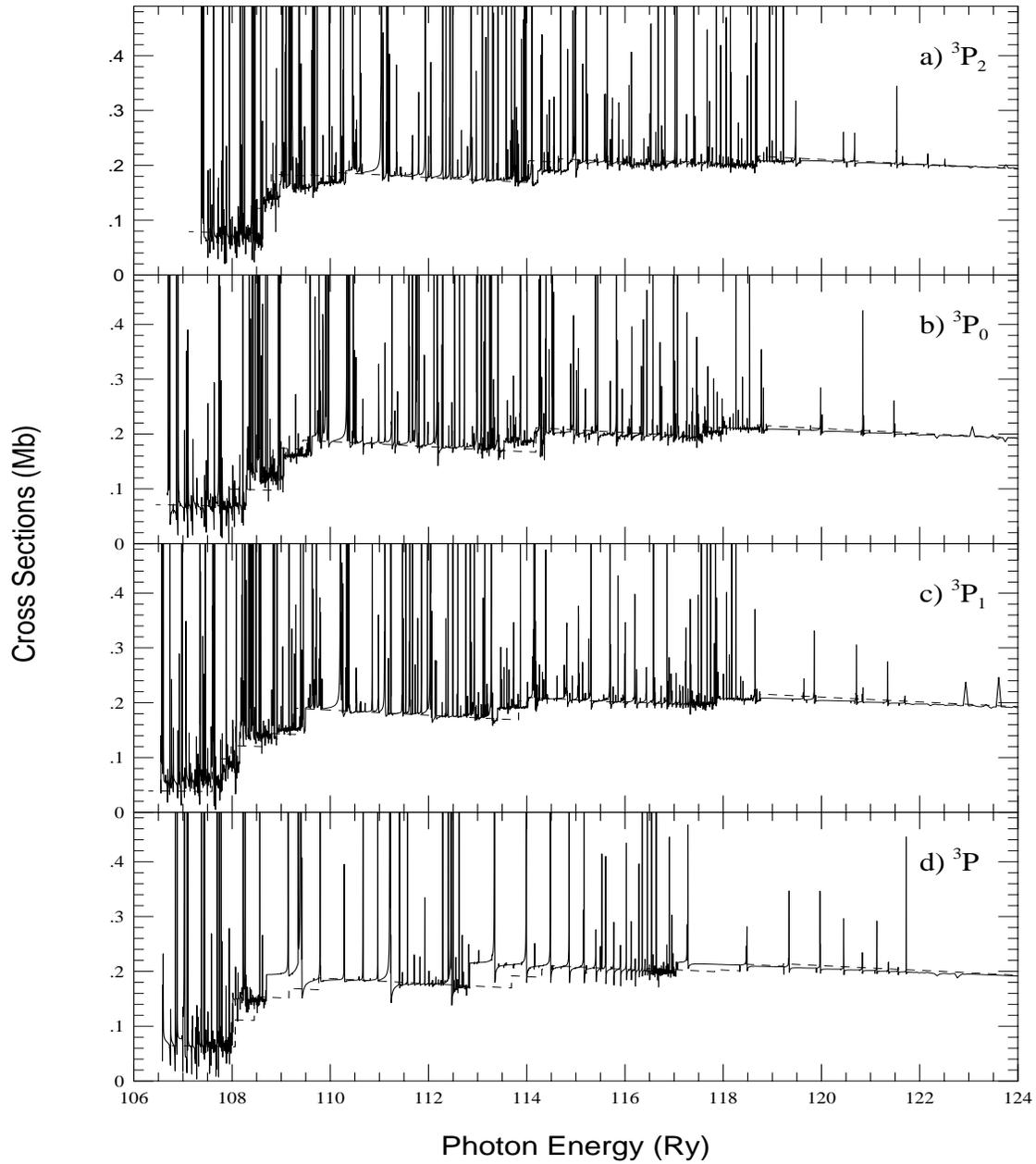,height=20.0cm,width=18.0cm}
\caption{As in Fig.~1, but with resonance structures at
high-resolution.}
\end{figure}

\clearpage
\begin{figure}
\centering
\psfig{figure=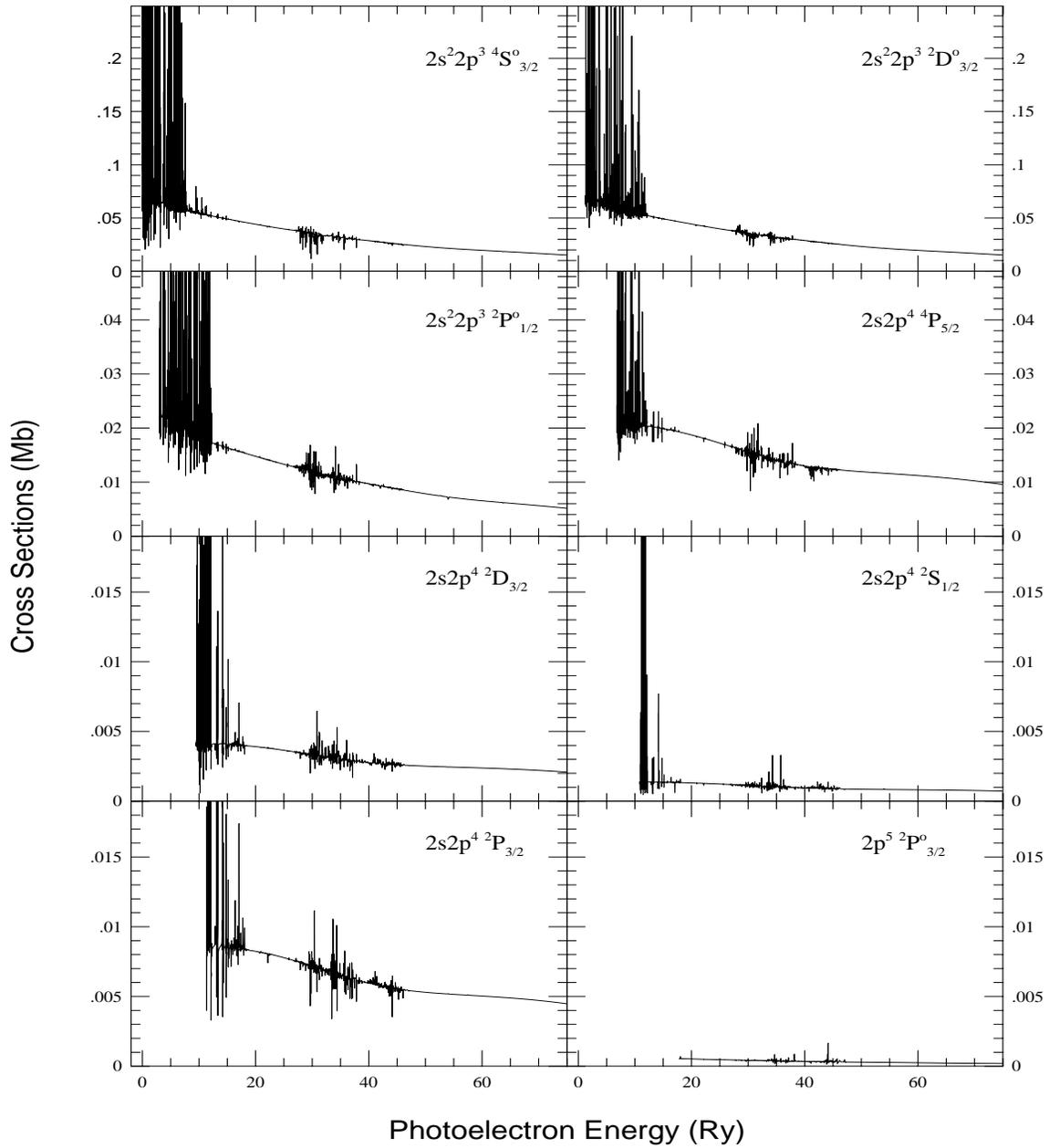,height=20.0cm,width=18.0cm}
\caption{Partial photoionization cross sections for the transitions
from the ground level $2s^22p^4\ ^3P_2$ in Fe~XIX to one of 
the low-lying levels in the residual ion Fe~XX by the present
BPRM calculation.}
\end{figure}

\label{lastpage}

\end{document}